\documentclass[a4paper]{jpconf}
\usepackage{graphicx,cite}
\begin{document}
\title{Emergence of magnetic long-range order in kagome quantum
antiferromagnets}

\author{Johannes Richter and Oliver G\"otze}

\address{Institut f\"ur Theoretische Physik, Universit\"at
Magdeburg,
39016 Magdeburg, Germany}

\ead{johannes.richter@physik.uni-magdeburg.de}

\begin{abstract}
The existence of a  spin-liquid ground state of the $s=1/2$ 
Heisenberg
kagome antiferromagnet (KAFM) is well established. Meanwhile, also for the $s=1$
Heisenberg KAFM evidence for the absence of magnetic long-range order (LRO) was found.
Magnetic LRO in
Heisenberg
KAFMs can emerge by increasing
the spin quantum number $s$ to $s>1$ and for $s=1$ by an easy-plane anisotropy.
In the present  paper we discuss the route to magnetic order in 
$s=1/2$ KAFMs by including an isotropic interlayer coupling (ILC) $J_\perp$ as well
as an easy-plane
anisotropy in the kagome layers by using the coupled-cluster
method to high orders of approximation.
We consider
ferro- as well as antiferromagnetic
$J_\perp$. 
To discuss 
the general
question for the crossover from a purely two-dimensional (2D) to a
quasi-2D and finally to a three-dimensional system we  consider
the simplest model of stacked (unshifted) kagome layers.
Although
the ILC of real kagome compounds is
often   
more sophisticated,
such a geometry of the ILC can be relevant for barlowite.
We find 
that the spin-liquid ground state  present for the strictly 2D $s=1/2$ $XXZ$
KAFM
survives a finite 
ILC, where the spin-liquid region shrinks monotonously with increasing
anisotropy.
If the ILC becomes large enough (about 15\% of 
intralayer coupling for the isotropic Heisenberg case
and about 4\% for the $XY$ limit) 
magnetic LRO  can be established, where the $q=0$ symmetry is favorable if
$J_\perp$ is
of moderate strength. 
If the strength of the ILC further increases, $\sqrt{3}\times \sqrt{3}$
LRO can become favorable against $q=0$ LRO. 
\end{abstract}

\section{Introduction}
One of the most fascinating  problems  in frustrated quantum magnetism 
is the investigation of the ground state (GS) of the $s=1/2$ antiferromagnet
 on the kagome lattice, see,
e.g.,~Refs.~\cite{singh1992,Waldtmann1998,Capponi2004,Singh2007,Sindzingre2009,Evenbly2010,Yan2011,lauchli2011,lee2011,iqbal2011,nakano2011,goetze2011,schollwoeck2012,becca2013,ioannis2013,bruce2014,Ioannis2014,Vishwanath2015,Becca2015,Oitmaa2016}.
The existence of a  spin-liquid ground state of the $s=1/2$ 
Heisenberg kagome
antiferromagnet (KAFM) is well established, although the discussion of
the nature of the spin-liquid state is
still not settled. Meanwhile, also for the $s=1$
Heisenberg
KAFM evidence for the absence of magnetic long-range order (LRO) was
found\cite{goetze2011,
Lauchli_s1_2014,Weichselbaum_s1_2014,satoshi_s1_2014,Weichselbaum_s1_2014a}. 
On the other hand, recently it has been reported that magnetic
$\sqrt{3}\times \sqrt{3}$ LRO can be
established  by increasing
the spin quantum number to $s>1$ \cite{goetze2011,Oitmaa2016,Liu2016}.
Further increasing $s$ stabilizes this kind of order, thus a
coupled-cluster study reports for the  magnetic order parameter (sublattice
magnetization) $M/s=0.074$, $0.203$, $ 0.294$, and  $0.358$ for $s=3/2$,
$2$, $5/2$ and $3$,
respectively \cite{goetze2011}. Note that in \cite{Oitmaa2016} for $s=3/2$ a similar value for
$M$ was
reported, whereas in \cite{Liu2016} a siginifcantly
larger $M$ was obtained.

Another way to decrease  quantum fluctuations is to     
include exchange anisotropy \cite{cepas2008,Mila2009,zhito_XXZ_2014,XXZ_s12_2014,wir_XXZ_2015,XXZ_s12_2015,Becca2015,fradkin2015,Chernyshev2015,Jaubert2015,Liu2016}
or to take into account 
further-neighbor couplings
\cite{XXZ_s12_2015,Domenge2005,Janson2008,Bishop2010,tay2011,Li2012,Balents2012,Thomale2014,Trebst2014,Gong2015,Gong2015a,Schollwoeck2015,Bieri2015,Laeuchli2015,Thomale2015}.
Such modifications of the pure nearest-neighbor (NN)
Heisenberg KAFM may play a crucial role to establish GS magnetic LRO of
$\sqrt{3} \times \sqrt{3}$ or of $q=0$ symmetry.   
For the $XXZ$ $s=1/2$ KAFM with NN couplings
only, an easy-axis  anisotropy is not sufficient to establish magnetic LRO
\cite{XXZ_s12_2014,wir_XXZ_2015,XXZ_s12_2015,Becca2015,Jaubert2015}.  
On the other hand, for $s=1$ there is evidence \cite{zhito_XXZ_2014,wir_XXZ_2015,Oitmaa2016} that 
a finite region of disorder around the isotropic Heisenberg point gives way
for  $\sqrt{3} \times \sqrt{3}$ LRO as increasing the  easy-axis anisotropy.
Further increasing the anisotropy towards the $XY$ model yields a second
transition from  $\sqrt{3} \times \sqrt{3}$ to $q=0$ LRO.

An obvious route towards magnetic LRO in the GS and also at finite
temperatures is given by including
a coupling between kagome layers.
Here
we investigate the effect of an isotropic interlayer coupling (ILC) on the ground state
order in a stacked $s=1/2$ 
$XXZ$ KAFM with easy-plane anisotropy in the layers
by using the coupled-cluster
method (CCM) to high orders of approximation.
The corresponding Hamiltonian reads
\begin{eqnarray}
\label{ham}
H=\sum_n\Bigg \{\sum_{\langle ij \rangle}\left ( s^x_{i,n} s^x_{j,n} + s^y_{i,n}
s^y_{j,n} + \Delta s^z_{i,n} s^z_{j,n}
 \right) \Bigg \} 
 + J_\perp \sum_{i,n} {\bf s}_{i,n} \cdot {\bf s}_{i,n+1},
\end{eqnarray}
where $n$ labels the kagome layers  
and  $J_\perp$ is a perpendicular (i.e., non-frustrated) ILC.
The expression in curly brackets
represents the 
$XXZ$ KAFM of the layer $n$ with NN intralayer couplings
$J=1$ and an easy-axis anisotropy $0 \le \Delta \le 1$.
We do not restrict the sign of the ILC, i.e.,
we consider ferro- as well as antiferromagnetic
$J_\perp$. 

It is in order to notice that the ILC in real kagome compounds is
often   
more sophisticated than that one considered here.
Hence, the aim of our paper is not the discussion of particular kagome
compounds, rather it is a theoretical investigation of the general
question for the crossover from a purely two-dimensional (2D) to a
quasi-2D and then to a three-dimensional system. 
For that it
is appropriate to consider
the simplest model of stacked unshifted kagome layers with a perpendicular
ILC.
There is at least one kagome compound,  namely barlowite,
with unshifted kagome
layers, see, e.g. \cite{barlowite2014,barlowite2014a}.
As it has been pointed out very recently \cite{barlow2016}, through isoelectronic substitution
in barlowite 
this kagome system becomes quite similar to our model system.

As already mentioned above, for isolated $s=1/2$ kagome layers the GS is always magnetically
disordered. 
Very recently, for the fully isotropic model it has been found that this
non-magnetic ground state persists until
relatively large strengths of the ILC \cite{our_EPL}, i.e., the critical
strength of the ILC $J^c_{\perp}$ needed to establish magnetic LRO is about 15\%
of the intralayer coupling.  
The question arises how an additional $XXZ$ anisotropy in the layers 
influences $J^c_{\perp}$.
Does the magnetically disordered GS survive also for the extreme limit of $XY$
layers, if the ILC becomes nonzero?
Bearing in mind that the GS selection for the strictly 2D model
depends on the anisotropy parameter $\Delta$ \cite{zhito_XXZ_2014,wir_XXZ_2015,Oitmaa2016}
we may also ask which GS symmetry ($\sqrt{3} \times \sqrt{3}$ or of $q=0$)
is selected when including the ILC.
Moreover, we will discuss to what extent the sign of $J_{\perp}$ is relevant.

\section{Brief outline of the coupled-cluster method (CCM)}  
The CCM is meanwhile a well-established and well-tested method in frustrated
quantum magnetism, see, e.g.
\cite{Bishop2010,goetze2011,wir_XXZ_2015,Li2012,Schm:2006,
darradi08,Zinke2008,farnell09,richter2010,farnell11,archi2014,bishop2014,gapj1j2_2015,jiang2015,Li2015,our_EPL}.
Hence, we want to illustrate here only some basic features of
the CCM. At that we follow
\cite{goetze2011,wir_XXZ_2015}, where the CCM was applied to the
2D KAFM, and \cite{our_EPL}, where the isotropic stacked $s=1/2$ KAFM was
studied by means of the CCM.
For more general information on the CCM, see,
\cite{roger90,bishop91a,zeng98,bishop98a,bishop00,bishop04}.
Note first that the CCM yields results directly for number of sites $N\to\infty$. 
The starting point of the CCM calculation 
is 
 a normalized reference  state
$|\Phi\rangle$. 
From a quasi-classical perspective  that is 
the stacked  coplanar $\sqrt{3}\times\sqrt{3}$ state or 
$q=0$ state (see, e.g.,
Refs.~\cite{zhito_XXZ_2014,wir_XXZ_2015,Harris1992,sachdev1992,chub92,henley1995}).
Then we perform a rotation of the local axes of each of 
the spins such that all spins in the reference state align along the
negative $z$ axis. 
Within this rotated local spin coordinate system 
we define a complete set of 
multispin
creation operators $C_I^+ \equiv (C^{-}_{I})^{\dagger}$ related to this reference
state:
$|{\Phi}\rangle = |\downarrow\downarrow\downarrow\cdots\rangle ; \mbox{ }
C_I^+ 
= { s}_{n}^+ \, , \, { s}_{n}^+{ s}_{m}^+ \, , \, { s}_{n}^+{ s}_{m}^+{
s}_{k}^+ \, , \, \ldots \; ,
$,
i.e., the spin operators entering the multispin
creation operators are defined 
in the local coordinate frames. The indices $n,m,k,\ldots$ denote arbitrary lattice
sites.
The ket 
and bra GS eigenvectors
$|\Psi\rangle$ 
and $\langle\tilde{\Psi}|$ 
of the spin system 
are given  by
$|\Psi\rangle=e^S|\Phi\rangle \; , \mbox{ } S=\sum_{I\neq 0}a_IC_I^+ \; ; \;
$
$\langle \tilde{ \Psi}|=\langle \Phi |\tilde{S}e^{-S} \; , \mbox{ } \tilde{S}=1+
\sum_{I\neq 0}\tilde{a}_IC_I^{-} .$
For the coefficients
$a_I$ and $\tilde{a}_I$  in the CCM correlation operators, $S$ and $\tilde{S}$,
the ket-state 
and bra-state
equations
$\langle\Phi|C_I^-e^{-S}He^S|\Phi\rangle = 0 \; , \; 
\langle\Phi|{\tilde S}e^{-S}[H, C_I^+]e^S|\Phi\rangle = 0 \;  , \; \forall
I\neq 0$, hold, where
each equation belongs to a certain multispin configuration $I$,
i.e., to a certain configuration of lattice sites
$n,m,k,\dots\;$.
For 
the GS energy $E_0=\langle\Phi|e^{-S}He^S|\Phi\rangle$ holds.
The   sublattice magnetization (magnetic order parameter) is given
by $ M = -\frac{1}{N} \sum_{i=1}^N \langle\tilde\Psi|{ s}_i^z|\Psi\rangle$, where
${s}_i^z$
is expressed in the rotated local coordinate system. 
For the many-body problem at hand,
we have to truncate the expansions of $S$ 
and $\tilde S$.
For that
we use the standard LSUB$m$ approximation
scheme, cf., e.g.,
\cite{roger90,bishop91a,Bishop2010,goetze2011,wir_XXZ_2015,Li2012,Schm:2006,
darradi08,Zinke2008,farnell09,richter2010,farnell11,bishop2014,gapj1j2_2015,jiang2015,Li2015,our_EPL},
where no more than $m$ spin flips spanning a range of no more than
$m$ contiguous lattice sites are included.
Using the 
parallelized CCM code \cite{cccm} we can solve the set of CCM equations up
to LSUB8. For this approximation level the number of CCM equations is
$278,024$.
We then extrapolate the `raw'
LSUB$m$
data to the limit $m \to \infty$. 
In correspondence to \cite{our_EPL} 
we use two schemes, namely an
extrapolation  using $m=4,5,\ldots,8$ (scheme I)
and separately an extrapolation  using $m=4,6,8$ (scheme II).
Scheme I corresponds to that one used for the 2D KAFM \cite{goetze2011,wir_XXZ_2015}, whereas scheme II
(not including the odd LSUB$m$ approximation levels) is more appropriate for
magnets with collinear AFM correlations 
\cite{bishop04,Schm:2006,darradi08,Zinke2008,richter2010,farnell11,archi2014,gapj1j2_2015,ccm_odd_even}.
By comparing the results of schemes I and II we can get
information  on the precision of the extrapolated
data. 

For the GS energy the formula 
$e_0(m)=E_0(m)/N = e_0(m\to\infty)  + a_1/m^2 + a_2/m^4$ is well tested and
it provides  
precise data for the extrapolated energy $ e_0(m\to\infty)$;
for the magnetic order parameter $M$ the ansatz
$M(m)=M(m\to\infty)+b_1(1/m)^{1/2}+b_2(1/m)^{3/2}$ is an appropriate choice to determine
quantum critical points
for frustrated quantum spin systems
\cite{Bishop2010,goetze2011,wir_XXZ_2015,Li2012,darradi08,Zinke2008,Schm:2006,richter2010,farnell11,gapj1j2_2015,Li2015,our_EPL}.

\begin{figure}[ht]
\includegraphics[width=19pc]{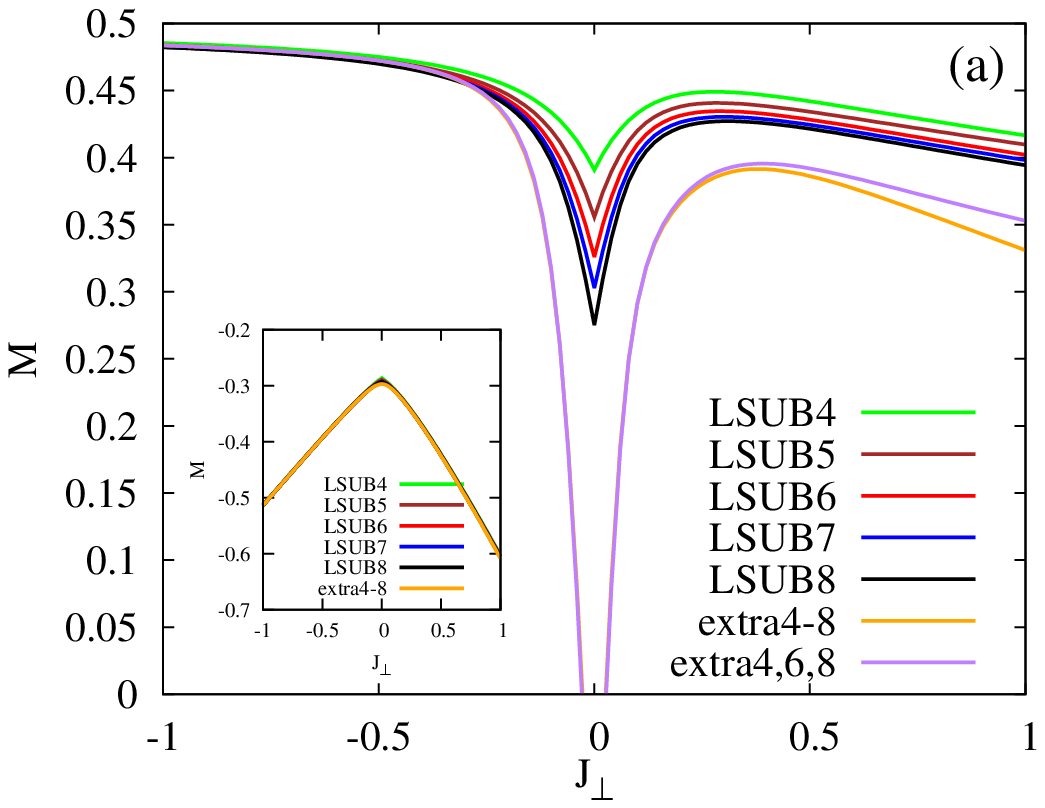}
\includegraphics[width=19pc]{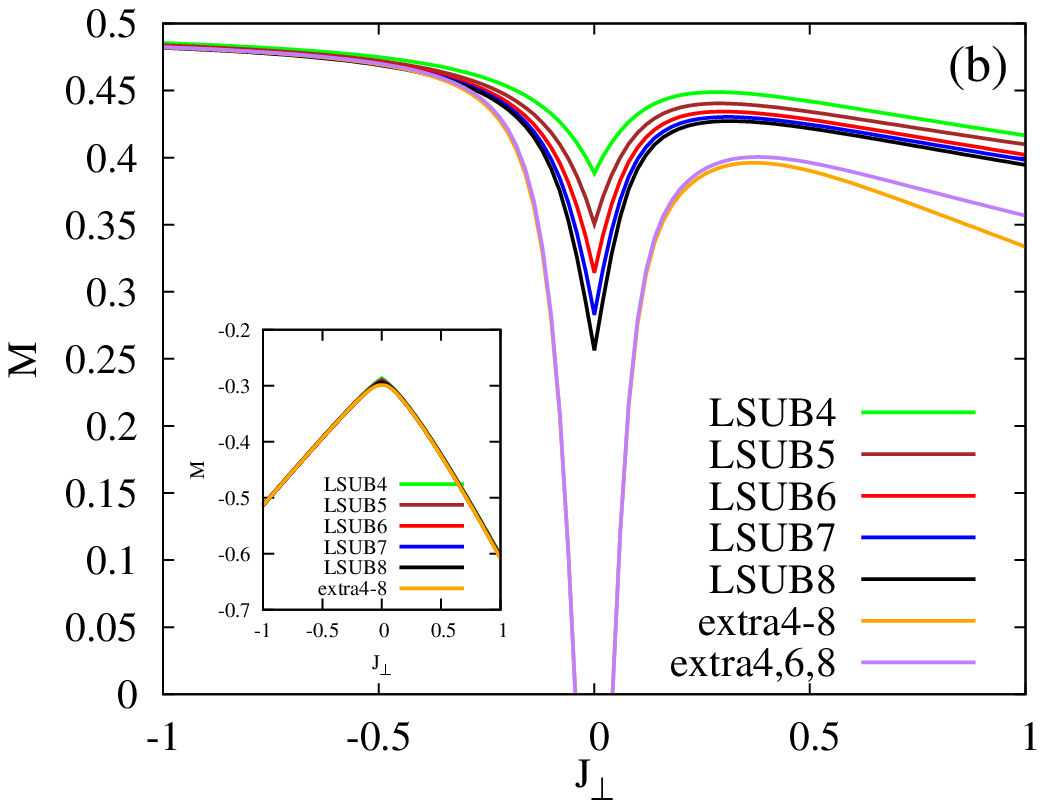}
\caption{\label{fig1}
CCM-LSUB$m$ as well as extrapolated GS sublattice magnetization $M$ for
$\Delta=0$  ($XY$ limit)
 using (a) the
$\sqrt{3}\times\sqrt{3}$ reference state  and (b) the $q=0$
 reference state 
as a function of the ILC $J_\perp$. 
The labels 'extra4-8' and 'extra4,6,8' correspond to the extrapolation
schemes I
and II, respectively (see main text).    
Inset: Corresponding data for the GS
energy $e_0$.   
Note that for $e_0$ the results   
of both
extrapolation schemes are practically identical, and, therefore we show only
extrapolated data for scheme I.}
\end{figure}

\section{Results}
We start with a discussion of the most anisotropic $XY$ limit, i.e.,
$\Delta=0$. 
 For that case we
present in Figs.~\ref{fig1}(a) and (b) the full set of LSUB$m$ data as well
as the extrapolated data for the order parameter $M$ (main panels) and also
for  the GS energy per spin $e_0=E_0/N$ (insets)
for the
$\sqrt{3}\times\sqrt{3}$ and $q=0$ reference states, respectively.
(Note that corresponding
figures for $\Delta=1$ can be found in \cite{our_EPL}.)
It is evident that $e_0$ converges quickly as the level $m$ of the LSUB$m$
approximation increases.
As a result, the extrapolation with leading order $1/m^2$
is very accurate, as it has been demonstrated in many
cases, where data from other precise methods are available to compare with,
see, e.g., Refs.~\cite{bishop04,goetze2011,wir_XXZ_2015}.
Obviously, the shape of the curves and the magnitude of the energies are very similar
for
both states, where from \cite{wir_XXZ_2015} it is known that for the 2D $XY$
model, i.e.,  at
$J_\perp=0$ and $\Delta=0$, the $q=0$ state has slightly lower energy. 
As found previously
\cite{XXZ_s12_2014,wir_XXZ_2015,XXZ_s12_2015,Becca2015,Jaubert2015},
$M$ is zero for $J_\perp=0$. 
For $\Delta=0$ the critical ILCs, where magnetic LRO sets in, are 
: $J^c_\perp=-0.027$, $J^c_\perp=+0.026$ ($\sqrt{3}\times\sqrt{3}$
state) and  $J^c_\perp=-0.044$, $J^c_\perp=+0.042$
($q=0$  state) when using scheme I,
and, $J^c_\perp=-0.028$, $J^c_\perp=+0.028$ ($\sqrt{3}\times\sqrt{3}$
state) and  $J^c_\perp=-0.044$, $J^c_\perp=+0.042$ ($q=0$ state)  when using scheme II.
Corresponding values for $\Delta=1$ are \cite{our_EPL}: 
$J^c_\perp=-0.100$, $J^c_\perp=+0.102$ ($\sqrt{3}\times\sqrt{3}$
state) and  $J^c_\perp=-0.154$, $J^c_\perp=+0.151$ ($q=0$  state) for scheme I,
and, $J^c_\perp=-0.104$, $J^c_\perp=+0.110$ ($\sqrt{3}\times\sqrt{3}$
state) and  $J^c_\perp=-0.135$, $J^c_\perp=+0.130$ ($q=0$ state)  for scheme II.
We may conclude that (i)
there is  a good agreement of the critical ILCs obtained by both 
extrapolation schemes, (ii) the spin-liquid region survives a finite ILC even in
the  $XY$ limit, and (iii) due to anisotropy this region becomes noticeably
smaller.

\begin{figure}[ht]
\includegraphics[width=19pc]{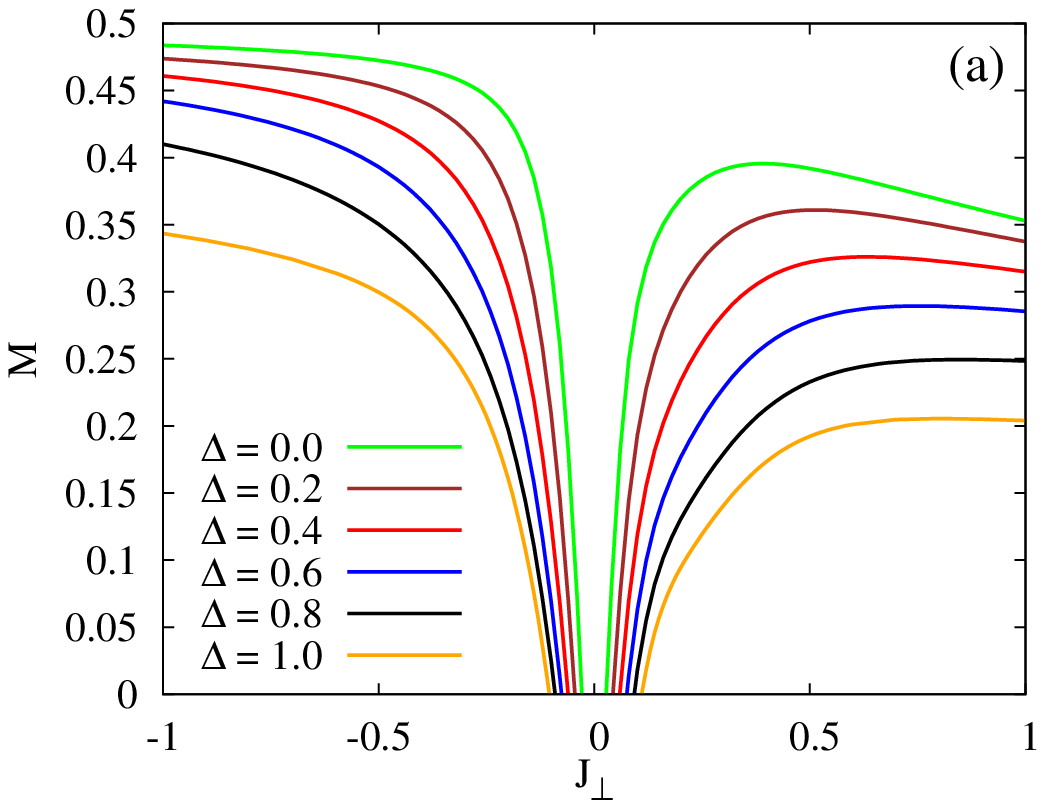}
\includegraphics[width=19pc]{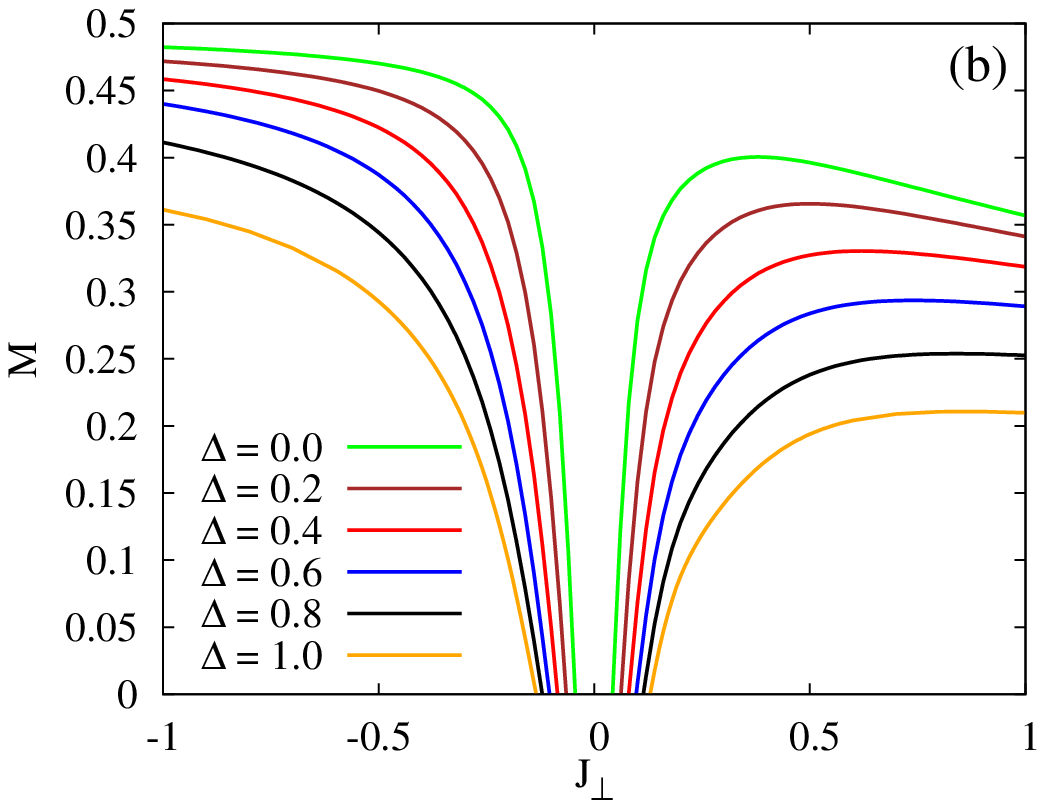}
\caption{\label{fig2} 
Extrapolated GS sublattice magnetization  $M$ (scheme II) for
various values of the anisotropy parameter $\Delta$  
 using (a) the
$\sqrt{3}\times\sqrt{3}$ reference state  and (b) the $q=0$
 reference state 
as a function of
the ILC $J_\perp$. 
}
\end{figure}

\begin{table}[ht]
\caption{\label{tab1}
Critical values  $J^c_\perp$ of the ILC where magnetic LRO
emerges for antiferromagnetic $J_\perp$ in dependence on the anisotropy
parameter $\Delta$.
The labels 'extra4-8' and 'extra4,6,8' correspond to the extrapolation
schemes I
and II, respectively (see main text).   
}  
\begin{center}
\lineup
\begin{tabular}{lllll}
\br
&\multicolumn{2}{c}{$q=0$}&\multicolumn{2}{c}{$\sqrt{3}\times\sqrt{3}$}\\
$\Delta$&extra4,6,8&extra4-8&extra4,6,8&extra4-8\\
\mr
$0.0$&$0.042$&$0.042$&$0.028$&$0.026$\\
$0.1$&$0.052$&$0.052$&$0.035$&$0.032$\\
$0.2$&$0.062$&$0.062$&$0.043$&$0.039$\\
$0.3$&$0.071$&$0.073$&$0.052$&$0.046$\\
$0.4$&$0.080$&$0.083$&$0.059$&$0.053$\\
$0.5$&$0.089$&$0.094$&$0.068$&$0.060$\\
$0.6$&$0.097$&$0.104$&$0.076$&$0.068$\\
$0.7$&$0.106$&$0.115$&$0.084$&$0.076$\\
$0.8$&$0.114$&$0.127$&$0.093$&$0.085$\\
$0.9$&$0.122$&$0.138$&$0.101$&$0.094$\\
$1.0$&$0.130$&$0.151$&$0.110$&$0.102$\\
\br
\end{tabular}
\end{center}
\end{table}

\begin{table}[ht]
\caption{\label{tab2}
Critical values $J^c_\perp$ of the ILC  where magnetic LRO
emerges for ferromagnetic $J_\perp$ in dependence on the anisotropy
parameter $\Delta$.
The labels 'extra4-8' and 'extra4,6,8' correspond to the extrapolation
schemes I
and II, respectively (see main text).    
}
\begin{center}
\lineup
\begin{tabular}{lllll}
\br
&\multicolumn{2}{c}{$q=0$}&\multicolumn{2}{c}{$\sqrt{3}\times\sqrt{3}$}\\
$\Delta$&extra4,6,8&extra4-8&extra4,6,8&extra4-8\\
\mr
$0.0$&$-0.044$&$-0.044$&$-0.029$&$-0.027$\\
$0.1$&$-0.055$&$-0.055$&$-0.037$&$-0.034$\\
$0.2$&$-0.065$&$-0.066$&$-0.045$&$-0.040$\\
$0.3$&$-0.075$&$-0.077$&$-0.053$&$-0.047$\\
$0.4$&$-0.085$&$-0.088$&$-0.061$&$-0.054$\\
$0.5$&$-0.094$&$-0.099$&$-0.069$&$-0.062$\\
$0.6$&$-0.104$&$-0.110$&$-0.076$&$-0.070$\\
$0.7$&$-0.112$&$-0.121$&$-0.083$&$-0.077$\\
$0.8$&$-0.120$&$-0.132$&$-0.091$&$-0.085$\\
$0.9$&$-0.128$&$-0.143$&$-0.098$&$-0.093$\\
$1.0$&$-0.135$&$-0.154$&$-0.104$&$-0.100$\\
\br
\end{tabular}
\end{center}
\end{table}

In the next figure we compare  the extrapolated magnetic order parameter $M$
for several values of the anisotropy parameter $\Delta$, see Figs.~\ref{fig2}
(a) and (b). Because (at least for small values of the $|J_\perp|$) both
extrapolation schemes lead to very similar results,  for convenience, we show only 
the data of scheme II. The critical values of the ILC, where magnetic LRO
sets in, are listed for both extrapolation schemes in Tables \ref{tab1} and \ref{tab2}.    
From  those tables and from Figs.~\ref{fig2} 
(a) and (b) we conclude:
(i) for all values $0 \le \Delta \le 1$ there is a finite region of magnetic
disorder,
(ii) this region shrinks monotonously with increasing anisotropy
(i.e. with  decreasing $\Delta$),    
(iii) near the critical ILC  $J^c_\perp$ the downturn of the magnetic
order parameter $M$ becomes very steep as increasing the anisotropy,
(iv) in the magnetically ordered phase the sublattice magnetization $M$ grows noticeably  as increasing
the anisotropy, and,
(v) while for ferromagnetic $J_{\perp} <0$  there is a monotonic increase of $M$ with
increasing $|J_\perp|$, for $J_\perp > 0$ we find a non-monotonic behavior
of $M$. Interestingly, for  smaller values of  $\Delta$ the maximum in $M$
appears at values of the ILC  significantly before 
$J_\perp=1$. For $\Delta =0$ it is at about $J_\perp = 0.36$
for both reference states.

\begin{figure}[ht]
\includegraphics[width=23pc]{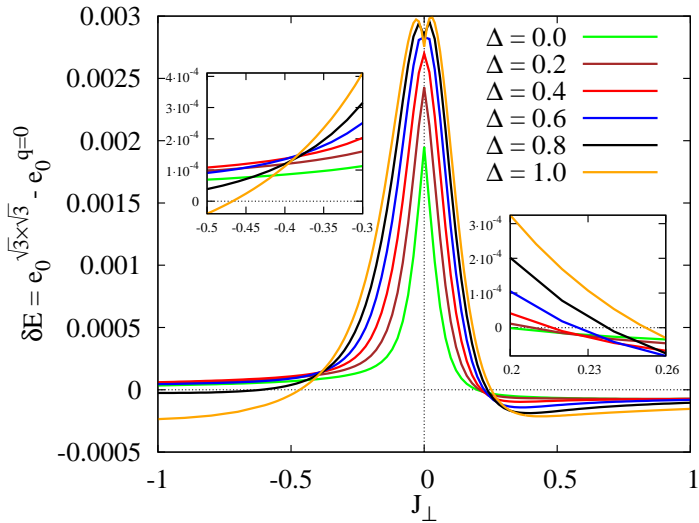}\hspace{2pc}%
\begin{minipage}[b]{13pc}\caption{\label{fig3}
Difference
$\delta e = e_0^{\sqrt{3} \times \sqrt{3}} - e_0^{q=0} $
of the extrapolated  GS energies (using scheme II) of the 
 $\sqrt{3}\times\sqrt{3}$ and the $q=0$ states 
as a function of the
ILC $J_\perp$. In the insets we show $\delta e $ in an enlarged scale for
those regions of $J_\perp$, where  $\delta e $ changes its sign.  
}
\end{minipage}
\end{figure}

\begin{figure}[ht]
\includegraphics[width=23pc]{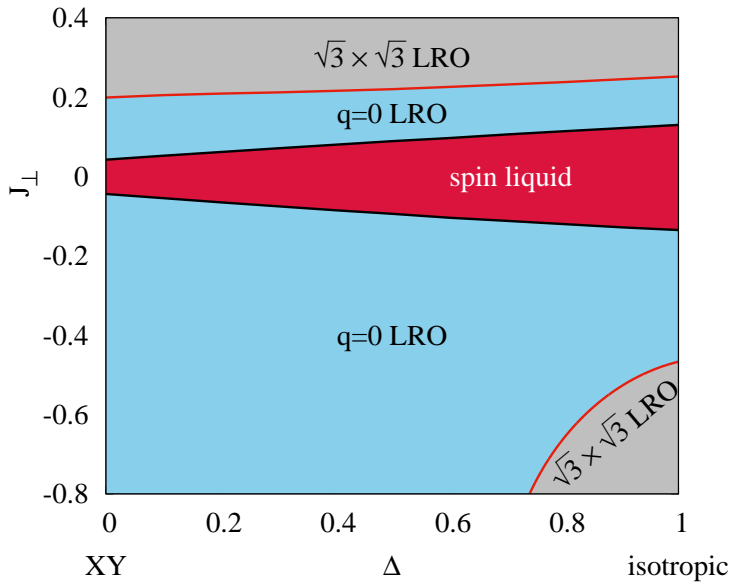}\hspace{2pc}%
\begin{minipage}[b]{13pc}\caption{\label{fig4}
Sketch
of a phase diagram
of the  stacked $s=1/2$
kagome antiferromagnet  with easy-plane anisotropy in the layers.
}
\end{minipage}
\end{figure}

Now we address the question which magnetic GS LRO is selected by quantum
fluctuations. As we know from previous studies, 
the
mechanism of quantum selection of the GS LRO in the KAFM is very subtle, and 
the energy difference between competing states 
can be very small \cite{zhito_XXZ_2014,wir_XXZ_2015,our_EPL}. 
The CCM approach  
in high orders of
approximation provides an accurate tool to compare these energies.
Thus, the CCM for $s>1/2$ yields the correct quantum selection of  the $\sqrt{3}\times\sqrt{3}$ GS vs. the
$q=0$ GS as 
obtained by non-linear spin-wave theory \cite{goetze2011}.
A direct comparison of CCM and
non-linear spin-wave data for energy differences (which are
also of the order $10^{-3} \ldots 10^{-4}$) for the $XXZ$ KAFM 
for large $s$, given in Fig.~3 of Ref.~\cite{wir_XXZ_2015},
demonstrates that both independent approaches agree very well.
In  Fig.~\ref{fig3} we show our results for the energy difference 
$\delta e = e_0^{\sqrt{3} \times \sqrt{3}}- e_0^{q=0} $ as a function of
$J_\perp$ for various values of $\Delta$. 
We mention first that both extrapolation
schemes I and II yield consistent results for $\delta e $. Therefore, 
 for convenience, we show in  Fig.~\ref{fig3} only
the data of scheme II.   
From Fig.~\ref{fig3} it is evident that 
at low values of  $|J_\perp|$ the $q=0$ reference state yields lower
energy, i.e. $\delta e > 0$. That is consistent  with
Refs.~\cite{goetze2011}
and \cite{wir_XXZ_2015}, where  the case $J_\perp=0$ was studied. 
For ferro- as well as antiferromagnetic ILC  $\delta e$ is still positive at those values of $J_\perp$,
where  the sublattice magnetizations $M_{\sqrt{3} \times \sqrt{3}}$ or
$M_{q=0}$ become larger than zero.
Hence, our results demonstrate that a magnetic disorder-to-order
transition to ${q=0}$  LRO takes place at $J^c_\perp$ as listed in 
columns 2 and 3 of Tables \ref{tab1} and \ref{tab2}.
We mention that the quantum selection of the $q=0$ GS LRO is contrary to the 
semi-classical large-$s$ selection of the $\sqrt{3} \times \sqrt{3}$ LRO 
found for the 2D spin-$s$ KAFM. 
As already found for the isotropic case,  $\Delta=1$, 
further increasing the strength of $J_\perp$ can lead to a second transition
at $J^t_\perp$ from  ${q=0}$  to $\sqrt{3} \times \sqrt{3}$ LRO determined
by the change of sign of $\delta e$.  
Here, we find such a second transition for all values  $\Delta$ on the
antiferromagnetic side, $J_\perp> 0$.  
On the ferromagnetic side a second transition is found only for $\Delta=1,
0.9, 0.8$.
For lower values of $\Delta $ the energy difference $\delta e$ remains
positive. However, it is in order to mention that already for $\Delta <
0.9$ our CCM result for $\delta e$ becomes extremely
small as $J_\perp \to -1$, and the ${q=0}$  and $\sqrt{3} \times \sqrt{3}$  states are practically
degenerate.  
We list or results for the position of the second transition between  ${q=0}$  and $\sqrt{3}
\times \sqrt{3}$ magnetic LRO in Table~\ref{tab3}.   
Summarizing our data for the transitions at $J^c_\perp$ and $J^t_\perp$  as
given in Tables~\ref{tab1}-\ref{tab3} we present in Fig.~\ref{fig4} a sketch of a phase diagram
of the  stacked $s=1/2$ 
$XXZ$ KAFM with easy-plane anisotropy in the layers.

\begin{table}[ht]
\caption{\label{tab3}
Transition points $J^t_\perp$ 
between  ${q=0}$  and $\sqrt{3}
\times \sqrt{3}$ magnetic LRO. For $\Delta <
0.75$ and ferromagnetic $J_\perp$ we do find a transition. }
\begin{center}
\lineup
\begin{tabular}{lllll}
\br
&\multicolumn{2}{c}{AFM}&\multicolumn{2}{c}{FM}\\
$\Delta$&extra4,6,8&extra4-8&extra4,6,8&extra4-8\\
\mr
$0.0$&$0.199$&$0.220$&\m---&\m---\\
$0.1$&$0.205$&$0.234$&\m---&\m---\\
$0.2$&$0.209$&$0.242$&\m---&\m---\\
$0.3$&$0.212$&$0.248$&\m---&\m---\\
$0.4$&$0.216$&$0.254$&\m---&\m---\\
$0.5$&$0.220$&$0.259$&\m---&\m---\\
$0.6$&$0.226$&$0.265$&\m---&\m---\\
$0.7$&$0.232$&$0.273$&\m---&\m---\\
$0.8$&$0.238$&$0.282$&$-0.614$&$-0.562$\\
$0.9$&$0.245$&$0.294$&$-0.502$&$-0.466$\\
$1.0$&$0.252$&$0.310$&$-0.467$&$-0.435$\\
\br
\end{tabular}
\end{center}
\end{table}

\section{Concluding remarks}

A main finding of our paper is 
that the spin-liquid GS present for the strictly 2D $s=1/2$ $XXZ$ KAFM
survives a finite 
ILC even in the $XY$ limit of maximal easy-plane anisotropy.
If the ILC becomes large enough  (about 15\% of the
intralayer coupling in our model system for the isotropic Heisenberg case
and about 4.5\% for coupled $XY$ kagome layers) 
magnetic LRO can be established, where the $q=0$ symmetry is favorable, if  $J_\perp$ is
of moderate strength. 
The quantum  selection of the $\sqrt{3}\times\sqrt{3}$ magnetic LRO versus the
$q=0$ magnetic LRO 
found for larger values of antiferromagnetic $J_\perp$
and also for ferromagnetic $J_\perp$ near the isotropic limit is related to a very small energy
gain. Thus, in real compounds  even very small additional terms in the relevant spin Hamiltonian such as
further distance exchange couplings may therefore be more relevant.

Let us finally discuss the relevance of our results for experiments
on kagome compounds.
As it has been pointed out very recently, through
isoelectronic substitution
in barlowite this kagome compound is related  to our model system of stacked
(unshifted) kagome
layers\cite{barlow2016}.
The finding that the spin-liquid phase 
survives a finite   
ILC  is relevant for the  modified  
barlowite system, where an ILC of about 6-7\%  of  the 
intralayer coupling was predicted
\cite{barlow2016}.
The relation to  herbertsmithite, for which  an ILC of about 5\% was determined
\cite{janson_diss},
is only on a qualitative level, since here  the kagome
layers are shifted from layer to layer. 
The observation of $q=0$ magnetic order in 
Cs$_2$Cu$_3$SnF$_{12}$ reported in \cite{Tanaka2014,Tanaka2015} 
is also in  qualitative agreement with our findings for the GS selection.
We also mention that anisotropic kagome antiferromagnets 
have relevance for the experimental research, see, e.g.,
Refs.~\cite{DM1,DM2,DM3,Tanaka2014}. 
Moreover, anisotropic spin models are of great interest with respect to 
engineering models of quantum magnetism on optical lattices, see, e.g.,
\cite{isakov2006,bloch2008,struck2013}.


\section*{References}
\begin{thebibliography}{9}


\bibitem{singh1992}R. R. P. Singh and D. A. Huse,
{\it Phys. Rev. Lett.} {\bf 68}, 1766 (1992).



\bibitem{Waldtmann1998}
C. Waldtmann et al.,
{\it Eur. Phys. J. B} {\bf 2}, 501 (1998).


\bibitem{Capponi2004}
S. Capponi, A. L\"auchli, and M. Mambrini,
{\it Phys. Rev. B} {\bf 70}, 104424 (2004). 


\bibitem{Singh2007}
R. R. P. Singh and D. A. Huse, 
{\it Phys. Rev. B} {\bf 76}, 180407(R) (2007).


\bibitem{Sindzingre2009}
P. Sindzingre and C. Lhuillier,
{\it Europhys. Lett.} {\bf 88}, 27009 (2009).


\bibitem{Evenbly2010}
G. Evenbly and G. Vidal,
{\it Phys. Rev. Lett.} {\bf 104}, 187203 (2010).


\bibitem{Yan2011}
S.~Yan, D.~A.~Huse, and S.~R.~White,
{\it Science} {\bf 332},  1173 (2011).

\bibitem{nakano2011} 
H. Nakano and T. Sakai, 
{\it J. Phys. Soc. Jpn.} {\bf 80}, 053704 (2011).

\bibitem{lauchli2011} 
A. M. L\"auchli, J. Sudan, and E. S. S{\o}rensen, 
{\it Phys. Rev. B} {\bf 83}, 212401 (2011).

\bibitem{lee2011}
Y.-M. Lu, Y. Ran, and P. A. Lee,
{\it Phys. Rev. B} {\bf 83}, 224413 (2011).




\bibitem{iqbal2011} 
Y. Iqbal, F. Becca, and D. Poilblanc, 
{\it Phys. Rev. B} {\bf 84}, 020407(R) (2011).

\bibitem{goetze2011}     O. G\"otze et al.,
          {\it Phys. Rev. B} {\bf 84}, 224428 (2011).

\bibitem{schollwoeck2012}
S. Depenbrock, I. P. McCulloch, and U. Schollw\"ock,
{\it Phys. Rev. Lett.} {\bf 109}, 067201 (2012).




\bibitem{becca2013}
Y. Iqbal, F. Becca, S. Sorella, D. Poilblanc,
{\it Phys. Rev. B} {\bf 87}, 060405(R) (2013).

\bibitem{ioannis2013} I. Rousochatzakis, R. Moessner, J. van den Brink,
 {\it Phys. Rev. B} {\bf 88}, 195109 (2013).




\bibitem{bruce2014} Z. Y. Xie et al.,
{\it Phys. Rev. X} {\bf 4}, 011025 (2014). 

\bibitem{Ioannis2014} I. Rousochatzakis, Y. Wan, O. Tchernyshyov, and F.
Mila,
{\it Phys. Rev. B} {\bf 90}, 100406(R) (2014).

\bibitem{Becca2015} W.-J. Hu, S.S. Gong, F. Becca, and D.N. Sheng,
 	{\it Phys. Rev. B} {\bf 92}, 2015(R) (2015).


\bibitem{Vishwanath2015}
M. P. Zaletel and A. Vishwanath,
{\it Phys. Rev. Lett.} {\bf 114}, 077201 (2015).


\bibitem{Oitmaa2016} J. Oitmaa and R. R. P. Singh,
{\it Phys. Rev. B} {\bf 93}, 014424 (2016)


\bibitem{Weichselbaum_s1_2014}
T. Liu, W. Li, A. Weichselbaum, J. von Delft, and G. Su,
{\it Phys. Rev. B} 91, {\bf 060403} (2015).

 

\bibitem{Lauchli_s1_2014}
H. J. Changlani and A. M. L\"auchli,
{\it Phys. Rev. B} {\bf 91}, 100407 (2015).




\bibitem{satoshi_s1_2014}
S. Nishimoto and M. Nakamura,
{\it Phys. Rev. B} {\bf 92}, 140412(R) (2015).


\bibitem{Weichselbaum_s1_2014a}
    W. Li, A. Weichselbaum, J. von Delft, and H.-H. Tu,
{\it Phys. Rev. B} {\bf 91}, 224414 (2015).


\bibitem{Liu2016} Tao Liu, Wei Li, Gang Su,
 arXiv:1603.01935v1.


\bibitem{cepas2008} O. Cepas, C. M. Fong, P. W. Leung, C. Lhuillier,
{\it Phys. Rev. B} {\bf 78}, 140405(R) (2008).

\bibitem{Mila2009} 
I. Rousochatzakis, S. R. Manmana, A. M. L\"auchli, B. Normand, and F. Mila,
{\it Phys. Rev. B} {\bf 79}, 214415 (2009).



\bibitem{zhito_XXZ_2014}  A. L. Chernyshev and M. E. Zhitomirsky,
        {\it Phys. Rev. Lett.} {\bf 113}, 237202 (2014).



\bibitem{XXZ_s12_2014}
Y.C. He and Y. Chen,  	{\it Phys. Rev. Lett.} {\bf 114}, 037201 (2015)



\bibitem{wir_XXZ_2015} O. G\"otze and J. Richter,
    {\it Phys. Rev. B} {\bf 91}, 104402 (2015).  


\bibitem{XXZ_s12_2015}
W. Zhu, S. S. Gong, and D. N. Sheng, 
 	{\it Phys. Rev. B} {\bf 92}, 014424 (2015).



\bibitem{fradkin2015}  K. Kumar, K. Sun, and E. Fradkin, 
 	{\it Phys. Rev. B} {\bf 92}, 094433 (2015).

\bibitem{Chernyshev2015} A. L. Chernyshev and M. E. Zhitomirsky,
{\it Phys. Rev. B} {\bf 92}, 144415 (2015).



\bibitem{Jaubert2015} K. Essafi, O. Benton, and L.D.C. Jaubert, 
 	{\it Nat. Commun.}  {\bf 7}, 10297 (2016)

\bibitem{Domenge2005} J.-C. Domenge, P. Sindzingre, C. Lhuillier, and 
L. Pierre,
{\it Phys. Rev. B} {\bf 72}, 024433 (2005).

\bibitem{Janson2008}
O.~Janson, J. Richter, and H.~Rosner,
       {\it Phys. Rev. Lett.} {\bf 101}, 106403 (2008).

\bibitem{Bishop2010}  R.F. Bishop, P.H.Y. Li, D.J.J. Farnell, and C.E. Campbell, 
{\it Phys. Rev. B} {\bf 82}, 104406 (2010).


\bibitem{tay2011} T. Tay and O. I. Motrunich,   {\it Phys. Rev. B} {\bf 84},
020404(R)
(2011).


\bibitem{Li2012} P.H.Y. Li, R.F. Bishop, C.E. Campbell, D.J.J. Farnell, O.~G\"otze, and J.~Richter,
       {\it Phys. Rev. B} {\bf 86}, 214403 (2012).



\bibitem{Balents2012}
H.-C. Jiang, Z. Wang, and L. Balents,
        {\it Nature Physics} {\bf 8}, 902 (2012).

\bibitem{Trebst2014}
B. Bauer et al.,
{\it Nature Communications} {\bf 5}, 5137 (2014).


\bibitem{Thomale2014}
    R. Suttner, C. Platt, J. Reuther, and R. Thomale,
{\it Phys. Rev. B} {\bf 89}, 020408(R) (2014).


\bibitem{Gong2015} W.J. Hu, W. Zhu, Y. Zhang, S.S. Gong, F. Becca, and D.N. Sheng
{\it Phys. Rev. B} {\bf 91}, 041124(R) (2015)

\bibitem{Gong2015a}
S.S. Gong, W. Zhu, L. Balents, and  D. N. Sheng
{\it Phys. Rev. B} {\bf 91}, 075112 (2015).


\bibitem{Schollwoeck2015} F. Kolley, S. Depenbrock, I. P. McCulloch,
U. Schollw\"ock, and V. Alba, 
{\it Phys. Rev. B} {\bf 91}, 104418 (2015).





\bibitem{Bieri2015}
S. Bieri, L. Messio, B. Bernu, and C. Lhuillier,
{\it Phys. Rev. B} {\bf 92}, 060407(R) (2015).

\bibitem{Laeuchli2015}    A. Wietek, A.Sterdyniak, and A. M.
 Laeuchli, {\it Phys. Rev. B} {\bf 92}, 125122 (2015).




\bibitem{Thomale2015}     Y. Iqbal et al.,
{\it Phys. Rev. B} {\bf 92}, 220404 (2015).


\bibitem{barlowite2014}  T.-H. Han, J. Singleton, J. A. Schlueter,
 {\it Phys. Rev. Lett.} {\bf 113}, 227203 (2014).

\bibitem{barlowite2014a} 
H. O. Jeschke et al.,
{\it Phys. Rev. B} {\bf 92}, 094417 (2015).

\bibitem{barlow2016}
D. Guterding, R. Valenti, and H. O. Jeschke, arXiv:1605.08162.

\bibitem{our_EPL}
O. G\"otze and J. Richter,
{\it Europhys. Lett. (EPL)} {\bf 114}, 67004 (2016).




\bibitem{Schm:2006}
D. Schmalfu{\ss}, R. Darradi, J. Richter, J. Schulenburg, and D. Ihle,
{\it Phys. Rev. Lett.} {\bf 97}, 157201 (2006).





\bibitem{darradi08}
R. Darradi, O. Derzhko, R. Zinke, J. Schulenburg, S.E. Kr\"uger and J. Richter,
{\it Phys. Rev. B} {\bf 78}, 214415 (2008).


\bibitem{Zinke2008}
R.~Zinke,  J.~Schulenburg, and J.~Richter,
      {\it Eur. Phys. J. B} {\bf 61}, 147 (2008).


\bibitem{farnell09}
D.J.J.~Farnell, R.~Zinke,  J.~Schulenburg, and J.~Richter,
{\it J. Phys.: Cond. Matter} {\bf 21}, 406002 (2009).


\bibitem{richter2010}  
J. Richter, R. Darradi, J.~Schulenburg, D.J.J. Farnell, and H.
Rosner,
{\it Phys. Rev. B} {\bf 81}, 174429 (2010).


\bibitem{farnell11} 
D.J.J. Farnell et al.,
{\it Phys. Rev. B} {\bf 84}, 012403 (2011).






\bibitem{archi2014}  D.J.J. Farnell et al.,
       {\it Phys. Rev. B} {\bf 89}, 184407 (2014).




\bibitem{bishop2014} 
R. F. Bishop, P. H. Y. Li, and C. E. Campbell,
{\it Phys. Rev. B} {\bf 89},   214413 (2014).



\bibitem{jiang2015} J.-J. Jiang, Y.-J Liu, F. Tang, C.-H. Yang, and Y.-B. Sheng,
{\it Physica B: Cond. Mat.} {\bf 463}, 30 (2015).







\bibitem{gapj1j2_2015}
J. Richter, R. Zinke, D.J.J. Farnell,
{\it Eur. Phys. J. B} {\bf 88}, 2 (2015). 


\bibitem{Li2015}
P. H. Y. Li, R. F. Bishop, and  C. E. Campbell,
{\it Phys. Rev. B} {\bf 91},   014426 (2015).

\bibitem{bishop98a} 
R.F. Bishop, in {\it Microscopic Quantum Many-Body
Theories and Their Applications}, edited by 
J. Navarro and A. Polls, {\it Lecture Notes in Physics} 
{\bf 510} (Springer, Berlin, 1998), p.1.

\bibitem{bishop04}
D.J.J.~Farnell and R.F.~Bishop,
in {\it Quantum Magnetism}, {\it Lecture Notes in Physics} {\bf 645}, 307 (2004)

\bibitem{bishop00} 
R.F.~Bishop, D.J.J.~Farnell, S.E.~Kr\"uger,  J.B.~Parkinson,
{\it J. Phys.: Condens. Matter} { \bf 12}, 6887 (2000).



\bibitem{roger90} M. Roger and J.H. Hetherington,
                {\it Phys. Rev. B} {\bf 41}, 200 (1990).

\bibitem{bishop91a} R.F. Bishop, J.B. Parkinson, and Y. Xian,
                  { {\it Phys. Rev. B}} {\bf 43}, R13782 (1991);
                  { {\it Phys. Rev. B}} {\bf 44}, 9425 (1991).


\bibitem{zeng98}
C.~Zeng, D.J.J.~Farnell, and R.F.~Bishop,
{\it J. Stat. Phys.} {\bf 90}, 327 (1998).


\bibitem{Harris1992}
A. B. Harris, C. Kallin and A. J. Berlinsky,
{\it Phys. Rev. B} {\bf 45}, 2899 (1992).
%

\bibitem{sachdev1992}
S. Sachdev,
{\it Phys. Rev. B} {\bf 45}, 12377 (1992).

\bibitem{chub92} A. Chubukov,
{\it Phys. Rev. Lett.} {\bf 69}, 832 (1992).



\bibitem{henley1995} 
C. L. Henley and E. P. Chan, 
{\it J. Magn. Magn. Mater.} {\bf 140-144}, 1693 (1995).


\bibitem{cccm} 
For the numerical calculation we use the program package `The
crystallographic CCM' (D.J.J. Farnell and J. Schulenburg).

\bibitem{ccm_odd_even}
D. J. J. Farnell and  R. F. Bishop,
{\it Int. J. Mod. Phys. B} {\bf 22}, 3369 (2008).


\bibitem{janson_diss}
O. Janson, {\it DFT based microscopic magnetic modeling for
low-dimensional spin systems}, Ph.D. thesis, Technische
Universit{\"{a}}t Dresden (2012)

\bibitem{Tanaka2014}
T. Ono, K. Matan, Y. Nambu, T. J. Sato, K. Katayama,
S. Hirata, and H. Tanaka, J. Phys. Soc. Jpn. {\bf 83}, 043701 (2014);
K. Katayama, N. Kurita, H. Tanaka,
arXiv:1412.5770.


\bibitem{Tanaka2015}
K. Katayama, N. Kurita, and H. Tanaka,
{\it Phys. Rev. B} {\bf 91}, 214429 (2015).


\bibitem{DM1}
K. Matan, D. Grohol, D. G. Nocera, T. Yildirim, A. B. Harris, S. H. Lee, S.
E. Nagler, and Y. S. Lee,
Phys. Rev. Lett. {\bf 96}, 247201 (2006).

\bibitem{DM2}
A. Zorko, S. Nellutla, J. van Tol, L. C. Brunel, F. Bert, F. Duc, J.-C.
Trombe, M. A. de Vries, A. Harrison, and P. Mendels,
Phys. Rev. Lett. {\bf 101}, 026405 (2008).

\bibitem{DM3}
A. Zorko, F. Bert, A. Ozarowski, J. van Tol, D. Boldrin, A. S. Wills, and P.
Mendels,
Phys. Rev. B {\bf 88}, 144419 (2013).


\bibitem{isakov2006}
    S. V. Isakov, S. Wessel, R. G. Melko, K. Sengupta, Yong Baek Kim,
Phys. Rev. Lett. {\bf 97}, 147202 (2006).

\bibitem{bloch2008}
I. Bloch, J. Dalibard, and W. Zwerger,
Rev. Mod. Phys. {\bf 80}. 885 (2008).

\bibitem{struck2013}
J. Struck, M. Weinberg, C. \"Olschl\"ager, P. Windpassinger, J. Simonet, K. Sengstock,
R. H\"oppner, P. Hauke,  A. Eckardt, M. Lewenstein, and L. Mathey,	
Nature Physics  {\bf 9}, 738 (2013).
 


\end{thebibliography}
\end{document}